\documentclass[10pt,letterpaper,twocolumn]{article} 

\usepackage{ol2}
\usepackage[draft]{}
\usepackage{amsmath}
\usepackage{lipsum}
\usepackage{float}
\usepackage{txfonts}
\usepackage{graphicx}
\usepackage{mathptmx}
\usepackage{anyfontsize}
\usepackage{t1enc}
\usepackage{nicefrac}
\usepackage{color, hyperref}
\usepackage[squaren]{SIunits}
\newcommand{\be}{\begin{equation}}
\newcommand{\ee}{\end{equation}}
\newcommand{\ber}{\begin{eqnarray}}
\newcommand{\eer}{\end{eqnarray}}
\newcommand{\de}{\end{equation*}}
\newcommand{\cer}{\begin{eqnarray*}}
\newcommand{\der}{\end{eqnarray*}}

\begin{document}

\twocolumn[ 

\title{Higher order optical vortices and formation of speckles}


\author{Salla Gangi Reddy$^{1,*}$, Shashi Prabhakar$^1$, Ashok Kumar$^2$, J. Banerji$^1$, and R. P. Singh$^1$}

\address{
$^1$Physical Research Laboratory, Navarangpura, Ahmedabad, India-380 009. \\
$^2$Instituto de Fisica, Universidade de Sao Paulo, Sao Paulo, 66318, Brazil. \\
$^*$Corresponding author: sgreddy@prl.res.in
}

\begin{abstract} 
We have experimentally generated higher order optical vortices and scattered them through a ground glass plate that results in speckle formation. Intensity autocorrelation measurements of speckles show that their size decreases with increase in the order of the vortex. It implies increase in angular diameter of the vortices with their order. The characterization of vortices in terms of their annular bright ring also helps us to understand these observations. The results may find applications in stellar intensity interferometry and thermal ghost imaging. 

\end{abstract}


 ] 

Optical vortices are phase singularities or screw dislocations in the light field \cite{berry}. These vortices have helical wavefronts and the Poynting vector of such fields rotates around the propagation axis of the light. Due to such rotation of the Poynting vector, optical vortices carry an orbital angular momentum $m\hbar$ per photon, $m$ is called the azimuthal index or topological charge of the vortex. The spatial structure of the optical vortices looks like a ring with a dark core at the center \cite{grier, core}. Due to their peculiar properties, they are getting a lot of attention and finding applications in optical manipulation \cite{OAM}, astronomy \cite{thide} as well as quantum information and computation \cite{torner}. 

The scattering of optical vortices through a rotating ground glass plate (GGP) can be used to control the temporal intensity correlation  of scattered light \cite{ashok,ashok1}. It has been shown that the decay of correlation becomes sharper with increase in the order. We have also observed the revival of the dark core in scattered optical vortices at the far field intensity distribution and modelled them as partially coherent optical vortices \cite{gangi}. In this article, we have experimentally studied speckles \cite{speckle, dainty, goodmann1} generated from the scattering of optical vortices through a GGP. We have also analysed the intensity distribution of annular bright ring, a characteristic of optical vortices, which form these speckles.

\begin{figure}[h]
\begin{center}
\includegraphics[width=3.0in]{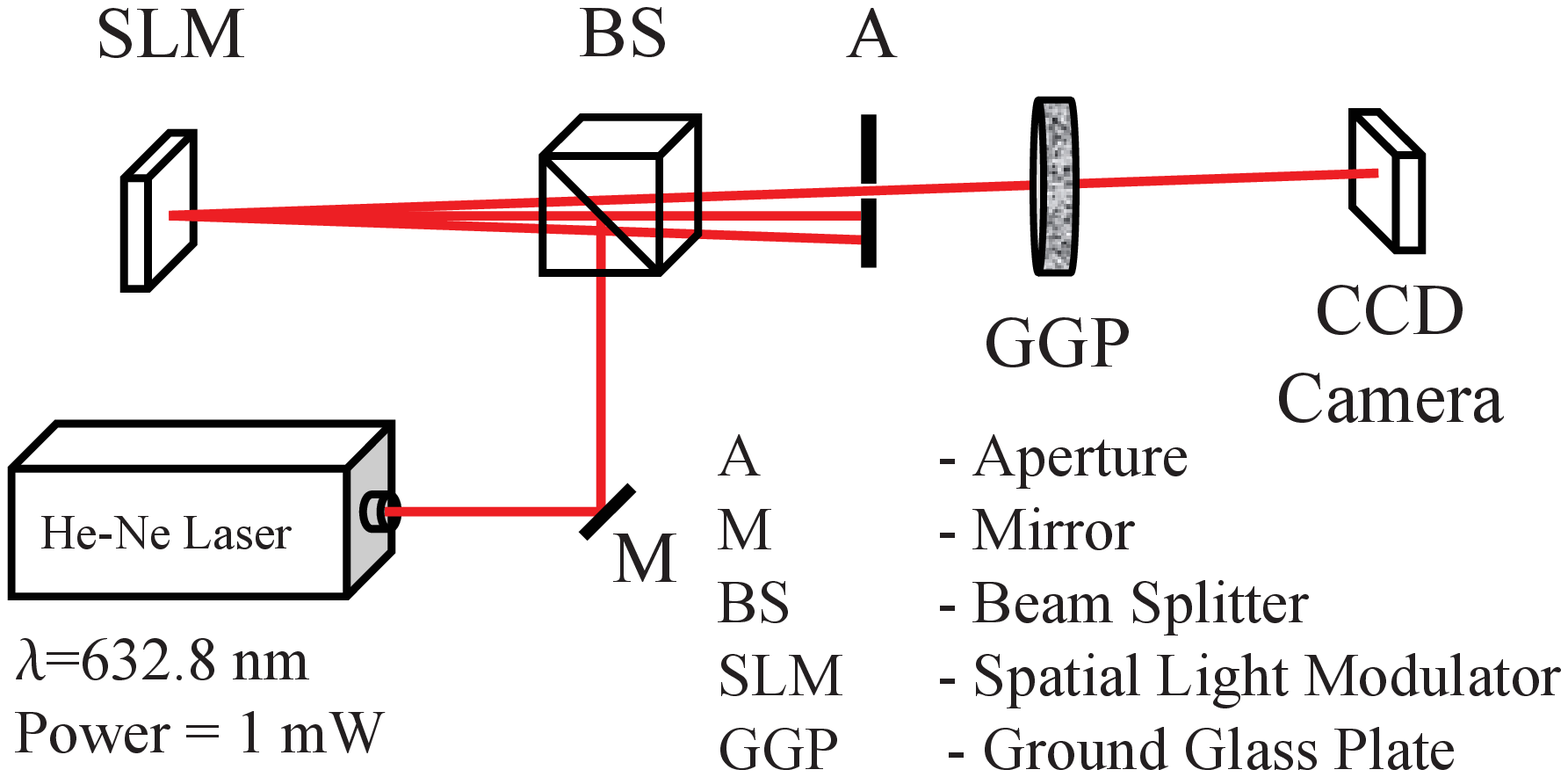}
\caption{ (Colour online) The experimental set-up for the generation and recording of speckles from optical vortices.}\label{fig:expt}
\end{center}
\end{figure}

\begin{figure}[h]
\begin{center}
\includegraphics[width=3.2in]{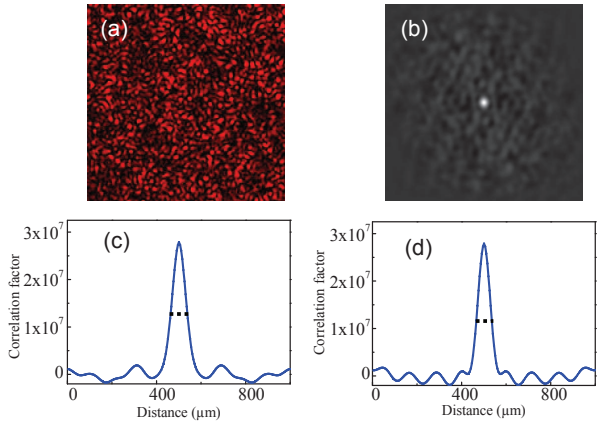}
\caption{ (Colour online) (a) The obtained speckle pattern of first order optical vortex, (b) the distribution of auto correlation function, (c) and (d) are the variation of autocorrelation function in both transverse directions X and Y respectively.} \label{fig:ac}
\end{center}
\end{figure}

Our experimental setup for the generation of speckles from optical vortices is shown in Fig. \ref{fig:expt}. An intensity stabilized He-Ne laser (Spectra Physics 117A) of power 1 mW and beam waist 0.3 mm is used to generate optical vortices. The optical vortex beams are produced by computer generated holography using a spatial light modulator (SLM) (Holoeye LCR 2500). Different computer generated holograms are introduced to the SLM through a computer for generating vortices of different order in the first diffraction order. The required beam is selected with an aperture $A$, and passed through the GGP. The scattered light from the GGP forms a granular pattern of intensity maxima and minima or speckles, as shown in Fig. \ref{fig:ac} (a). These speckles are recorded with a CCD camera (Evolution VF colour cooled). The SLM is placed at a distance of 60 cm from the laser and the GGP is at a distance of 66 cm from the SLM.

\begin{figure}[h]
\begin{center}
\includegraphics[width=3.2in]{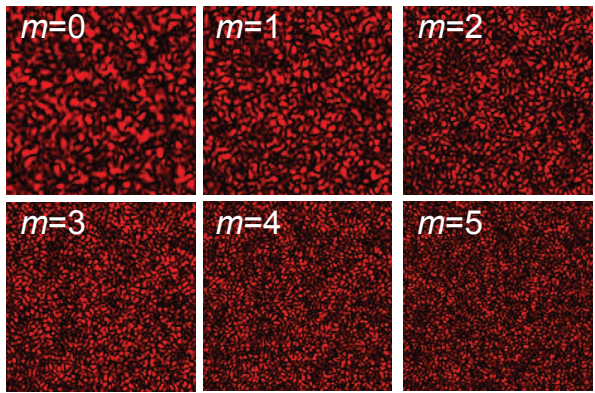}
\caption{ (Colour online) The speckle patterns formed by the vortices with orders $m$ = 0 -- 5 at a given plane.}\label{fig:speckle}
\end{center}
\end{figure}

The size of the recorded speckles has been determined by using auto-correlation method \cite{goodmann1} which calculates the correlation of speckle with itself. In this method, we fix one image of the speckles and observe its correlation numerically with a number of images shifted in position. These shifts can be made pixel by pixel in both the transverse directions. We plot the results as a function of the shift. The correlation factor is maximum if two speckle distributions are completely overlapped and it decreases with the decrease in overlap. The correlation factor becomes zero if their overlap is less than the speckle size due to the random nature of speckles. The correlation curve has a Gaussian distribution whose full width at half maximum (FWHM) gives the speckle size in any of the transverse directions. Here, we have considered the normalized speckle patterns to determine the speckle size as they are overfilling the CCD camera.

Fig. \ref{fig:ac} (a) shows the speckle pattern formed by the scattering of first order vortex through a GGP. The size of autocorrelation function is always twice the considered size of the speckle pattern as we are observing correlation along both the positive and negative directions. To find the size of recorded speckles, we have chosen $200 \times 200$ pixel sized speckle pattern and its autocorrelation as shown in Figs. \ref{fig:ac} (a) and \ref{fig:ac} (b) respectively. The distributions of the autocorrelation function in two transverse directions have been shown in Figs. \ref{fig:ac} (c) and \ref{fig:ac} (d). The FWHM of these two distributions provides the speckle size in the corresponding transverse directions.

\begin{figure}[h]
\begin{center}
\includegraphics[width=3.5in]{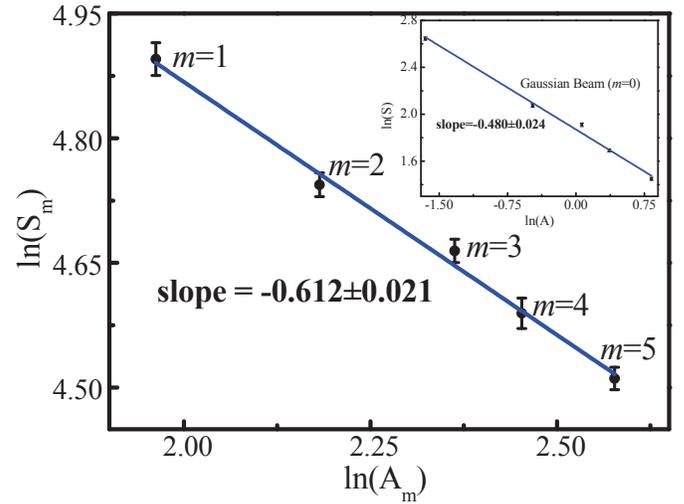}
\caption{ (Colour online) The plot of $\ln S$ versus $\ln A_m$ where $S$ and $A_m$ are the experimentally obtained speckle size and area of bright region of optical vortices respectively, (in inset, the same plot for a Gaussian beam with different areas). The solid line is the best fit for our experimental data.}\label{fig:mueller1}
\end{center}
\end{figure}

\begin{figure}[h]
\begin{center}
\includegraphics[width=3.0in]{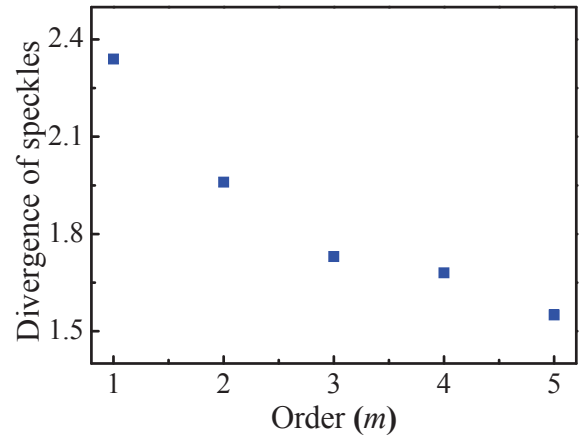}
\caption{ (Colour online) The plot of divergence of the speckle fields vs order of vortices.}\label{fig:mueller3}
\end{center}
\end{figure}


Fig. \ref{fig:speckle} shows the speckle patterns formed by the scattering of optical vortices of orders $m =$ 0 to 5 through the same GGP where $m=0$ corresponds to a Gaussian beam. These speckles have been recorded at a distance of 18 cm from the GGP. One can see very clearly that the speckle size $S_m$ decreases with the increase in order $m$. We have also shown the dependence of speckle size ($S_m$) on the area of the bright annular vortex ring ($A_m$). In Fig. \ref{fig:mueller1}, we show a plot of $ln(S_m)$ with $ln (A_m)$. From this graph also, it is clear that the speckle size decreases as the order of the vortex increases. The curve is a straight line with slope equal to the exponent of $A_m$. With the best fit to our experimental data, we have found that the speckle size is directly proportional to $A_{m}^{-0.612 \pm 0.021}$. In Fig. \ref{fig:mueller1}, we present the experimental data along with the best fit curve, which is different from the corresponding result for a Gaussian beam. For a Gausian beam scattered through a ground glass plate, one expects a Brownian distribution, and in fact one gets speckle size as proportional to $A^{-0.50}$ \cite{goodmann1, sirohi}. We have verified this experimentally by using different beam sizes of a Gaussian beam and shown in inset of Fig. \ref{fig:mueller1}. This suggests the non-Gaussian statistics of the speckles generated by the scattering of optical vortex beams. 

Also we study the rate of change of speckle size with propagation distance, namely divergence of speckles. For this, we have recorded speckles at seven different planes separated by a distance of 2cm from each and starting at 11cm from GGP and found their size for each order. The slope of a straight line obtained for speckle size vs propagation distance gives the divergence of speckles. We found that the divergence decreases with increase in the order as shown in Fig. \ref{fig:mueller3}.

We know that the size of speckles, the lowest length scale at which light is correlated, plays a crucial role in astronomy \cite{astro}. By finding the size of the speckles ($S$), one can determine the angular diameter ($W$) of the stars from the relation $ W = L\lambda/S $ where $L$ is the distance of the star from the observation plane and $\lambda$ is the wavelength. Taking an analogy, our experimental results show a decrease in speckle size with order implying an increase in the angular diameter of source generating these speckles i.e. optical vortices.

To analyse the change in speckle size with the order of vortex, we have analysed area of the annular bright ring that also varies with the order as indicated by our experimental observations in Fig. \ref{fig:mueller1}. It should be pointed out that the intensity distribution of optical vortices and the scaling of radius have also been discussed in ref. \cite{grier} to describe the motion of a trapped single particle around an optical vortex over a wide range of topological charges.

\begin{figure}[h]
\begin{center}
\includegraphics[width=3.5in]{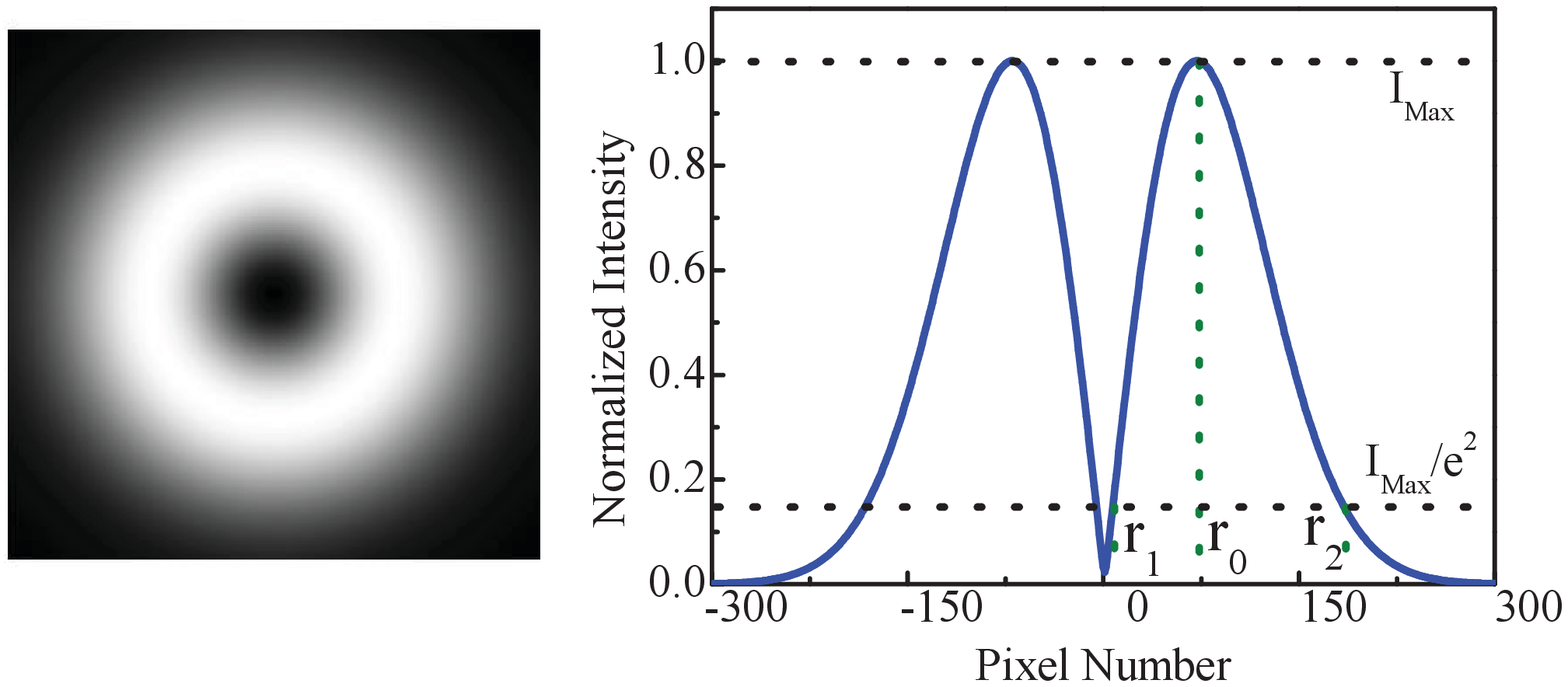}
\caption{ (Colour online) The transverse intensity distribution of an optical vortex of order 1 and its line profile.}\label{fig:vortex}
\end{center}
\end{figure}

\begin{figure}[h]
\begin{center}
\includegraphics[width=2.8in]{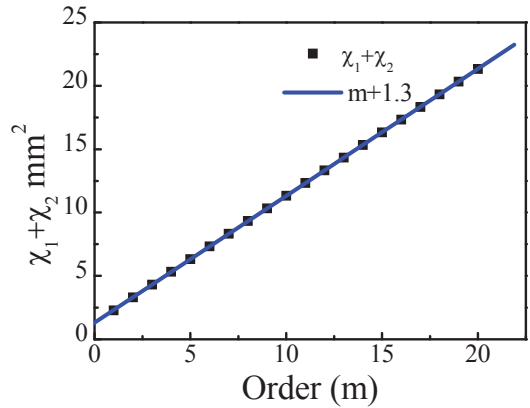}
\caption{ (colour online) Numerically obtained values of $\chi_2+\chi_1$ (filled squares) and the line $y=m+1.3$ as functions of $m$.}\label{fig:area}
\end{center}
\end{figure}

To find the area of annular bright ring, we start with the field distribution of a vortex of order $m$, embedded in a  Gaussian host beam of width $w_0$, as

\be
E_m(r)=(x+iy)^{m}\exp \left(-\frac{x^2+y^2}{w_0^{2}}\right)
\label{ov1}
\ee
and its intensity 
\be
I_m(r)=r^{2m}\exp \left(-\frac{2r^2}{w_0^{2}}\right),\qquad r^2=x^2+y^2.
\label{ov}
\ee
This intensity distribution is shown in Fig. \ref{fig:vortex}. Here, we are defining two parameters for a vortex beam: inner and outer radii ($r_{1}, r_{2}$). These are the radial distances at which the intensity falls to $1/e^2 (13.6\%)$ of the maximum intensity at $r=r_0$ (say). Here, $r_{1}$ is the point closer to the origin or the center and $r_{2}$ is the point farther from the center, the outer region of the beam.
The distances $r_i$ ($i=0,1,2$) can be obtained as follows. For the sake of convenience, we set $w_0=1$, that is, $w_0$ is the unit of measuring radial distances. We also define $\chi=r^2$ and $\chi_i=r_i^2$ ($i=0,1,2$) so that $I_m(r)=J_m(\chi)=\chi^m \exp(-2\chi)$. Differentiating $J_m(\chi)$ with respect to $\chi$, one easily obtains $\chi_0=m/2$ so that the maximum intensity has the value $J_m(\chi_0)=\chi_0^m \exp(-2\chi_0)$. The equations for $\chi_1$ and $\chi_2$ can then be written as
\begin{subequations}
\ber
\chi_1^m \exp(-2\chi_1)& = & \chi_0^m \exp(-2\chi_0-2)\\
\chi_2^m \exp(-2\chi_2)& = & \chi_0^m \exp(-2\chi_0-2).
\eer
\end{subequations}
These equations have been solved numerically for $m>0$. The numerical values are tabulated in Table \ref{khi1} and plotted in Fig. \ref{fig:area}.

\begin{table}[htb]
\begin{center}
\caption{Numerical solutions for Eqs. (2a) and (2b)}
\begin{tabular}{|l|l|l|l|}\hline
$m$ & $\chi_1$ & $\chi_2$ & $\chi_2+\chi_1$ \\ \hline
1 & 0.0262 & 2.2526 & 2.2789\\
2 & 0.1586  & 3.1462  & 3.3048\\
3 & 0.3602  & 3.9538  & 4.3140\\
4 & 0.6034  & 4.7154  & 5.3188\\
5 & 0.8748  & 5.4469  & 6.3216\\
6 & 1.1667  & 6.1569  & 7.3236\\
7 & 1.4746  & 6.8504  & 8.3250\\
8 & 1.7951  & 7.5309  & 9.3260\\
9 & 2.1262  & 8.2006  & 10.3268\\
10 & 2.4662  & 8.8613  & 11.3274\\
11 & 2.8138  & 9.5142  & 12.3280\\
12 & 3.1680  & 10.1605  & 13.3284\\
\dotfill & \dotfill &\dotfill &\dotfill \\
20 & 6.1682  & 15.1622  & 21.3304\\
50 & 18.5793  & 32.7529  & 51.3321\\
100 & 40.6553  & 60.6775  & 101.3330\\
200 & 86.5165  & 114.8165  & 201.3330\\
[1ex]\hline
\end{tabular}
\label{khi1}
\end{center}
\end{table}

Remarkably, it is found that to a very good approximation,
\be
\chi_2+\chi_1= m+1.3.
\ee
This empirical relationship can now be used to obtain simple expressions for $\chi_1$ and $\chi_2$. Multiplying the left sides of Eqs. (3a) and (3b), using Eq. (4) and the formula $(\chi_2+\chi_1)^2-(\chi_2-\chi_1)^2=4\chi_1\chi_2$, we get
\be
\chi_2-\chi_1=\sqrt{q_m},\qquad q_m=(m+1.3)^2-m^2\exp(-1.4/m).
\label{area}
\ee
Solving Eqs. (4) and (5), we get $\chi_1$, $\chi_2$ and hence, $r_1$, $r_2$:
\begin{subequations}
\ber
r_1& = &(m+1.3-\sqrt{q_m})^{1/2}/\sqrt{2}\\
r_2& = &(m+1.3+\sqrt{q_m})^{1/2}/\sqrt{2}.
\eer
\label{inner}
\end{subequations}

The area of the bright region in an optical vortex is given by 
\be
A_m=\pi (\chi_2-\chi_1)=\pi\sqrt{q_m}
\label{area2}
\ee
 which clearly depends on the order ($m$) of the vortex (see Eq. 5).

\begin{figure}[h]
\begin{center}
\includegraphics[width=3.0in]{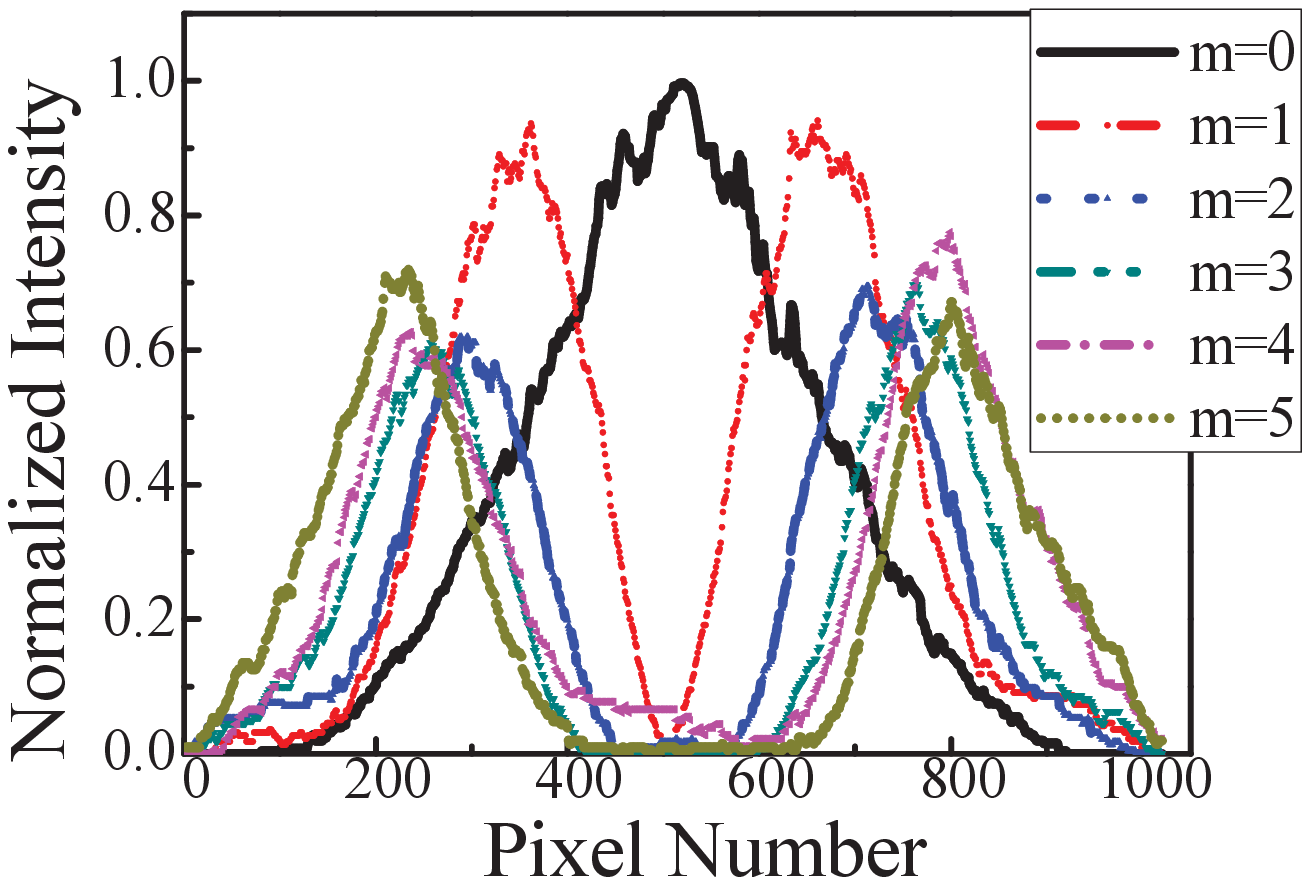}
\caption{ (Colour online) Line profiles along the vortex centres of optical vortices for orders 0 to 5 that are produced in the laboratory.}\label{fig:line}
\end{center}
\end{figure}

\begin{figure}[h]
\begin{center}
\includegraphics[width=3.0in]{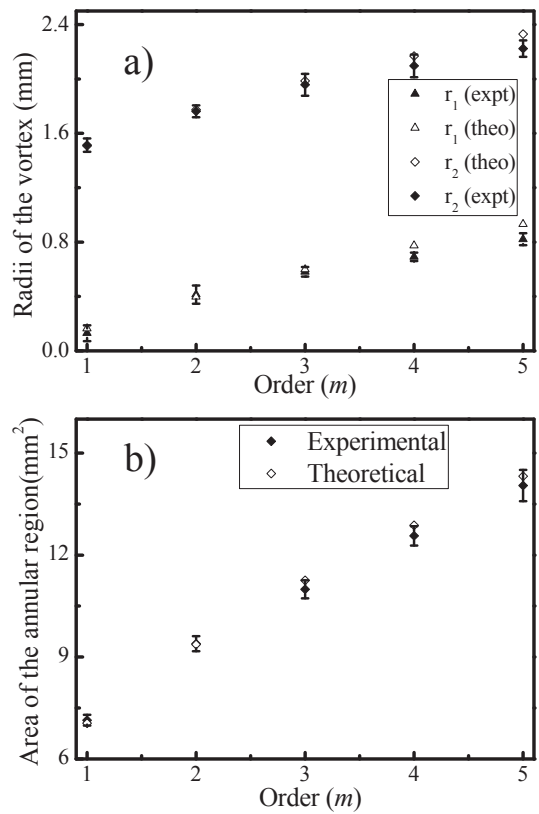}
\caption{ Experimental and theoretical results showing the variation of (a) inner and outer radii and (b) area of bright annular region with the order of a vortex.}\label{fig:area1}
\end{center}
\end{figure}

We also show experimentally the dependence of the area of the annular bright ring of optical vortices on their orders. Figure  \ref{fig:line} shows the line profiles of optical vortices for orders $m =$ 0 to 5 produced in the laboratory using computer generated holography technique. We have determined inner and outer radii of the vortex beams from the corresponding line profiles (Fig. \ref{fig:line}). The variation of inner and outer radii for vortex beams and the area of the annular bright ring with the order are shown in Fig. \ref{fig:area1}. The experimental findings are in good agreement with the theoretical values (obtained from Eqs. (6) and (7)) and prove that area of the annular bright ring increases in proportion to the order. We have calculated the inner and outer radii by taking the average over eight different line profiles of the experimentally obtained optical vortex. This can be correlated with decrease in size of the speckles with the order. 

It is known in the context of imaging that the spatial noise due to the speckles decreases if more number of speckles are present \cite{dainty}. We have shown that the speckle size decreases with the order of the vortex which effectively increases the number of speckles present in a given area. Therefore, one can use higher order vortices to reduce the spatial noise in speckle imaging. The results may find use in ghost imaging with vortices \cite{ghost1, ghost2, resolution} and in stellar intensity interferometry.

\pagebreak
\section*{Informational Fourth Page}


\begin{thebibliography}{99}

\bibitem{berry} J. F. Nye, and M. V. Berry, \emph{Dislocations in wave trains}, Proc. R. Soc. Lond. A Math. Phys. Sci. \textbf{336 (1605),} 165 (1974).

\bibitem{grier} J. E. Curtis, and D. G. Grier, \emph{Structure of Optical Vortices}, Phys. Rev. Lett. \textbf{90,} 133901 (2003).


\bibitem{core} A. Kumar, P. Vaity, Y. Krishna, and R. P. Singh, \emph{Engineering the size of dark core of an optical vortex}, Opt. Lasers Eng. \textbf{48,} 276 (2010).

\bibitem{OAM} L. Allen, S. M. Barnett, and M. J. Padgett, \textit{Optical Angular Momentum} (IOP, 2003).

\bibitem{thide} B. Thide, H. Then, J. Sjoholm, K. Palmer, J. Bergman, T. D. Carozzi, Y. N. Istomin, N. H. Ibragimov, and R. Khamitova, \emph{Utilization of Photon Orbital Angular Momentum in the Low-Frequency Radio Domain}, Phys. Rev. Lett. \textbf{99,} 087701 (2007).

\bibitem{torner} G. M. Terriza, J. P. Torres, and L. Torner, \emph{Twisted photons}, Nat. Phys. \textbf{3,} 305 (2007).

\bibitem{ashok} A. Kumar, J. Banerji, and R. P. Singh, \emph{Intensity correlation properties of high-order optical vortices passing
through a rotating ground-glass plate}, Opt. Lett. \textbf{35,} 3841 (2010).

\bibitem{ashok1} A. Kumar, J. Banerji, and R. P. Singh, \emph{Hanbury Brown -- Twiss -- type experiments with optical vortices and observation of modulated intensity correlation on scattering from rotating ground glass}, Phys. Rev. A \textbf{86,} 013825 (2012).

\bibitem{gangi} S. G. Reddy, A. Kumar, Shashi Prabhakar, and R. P. Singh, \emph{Experimental generation of ring-shaped beams with random sources}, Opt. Lett. \textbf{38,} 4441 (2013).

\bibitem{speckle} J. D. Rigden, and E. I. Gordan, \emph{The granularity of scattered optical maser light}, Proc. IRE, \textbf{50,} 2367 (1962).

\bibitem{dainty} J. C. Dainty, \textit{ Laser Speckle and related phenomena} (Springer Verlag, 1984).

\bibitem{goodmann1} J. W. Goodman, \emph{Speckle phenomena in optics,} (Indian edition 2008).

\bibitem{sirohi} R. Sirohi, \emph{Optical Methods of Measurements,} (Marcel Dekker, 1999).

\bibitem{astro} R. H. Brown and R. Q. Twiss, \emph{Quantum correlations in light}, Nature \textbf{177,} 27 (1956).

\bibitem{ghost1} S Crosby, S. Castelletto, C. Aruldoss, R. E. Scholten, and A Roberts, \emph{Modelling of classical ghost images obtained using scattered light}, New J. Phys. \textbf{9,} 285 (2007).

\bibitem{ghost2} M. N. O'Sullivan, K. W. C. Chan, and R. W. Boyd, \emph{Comparison of the signal-to-noise characteristics of quantum versus thermal ghost imaging}, Phys. Rev. A \textbf{82,} 053803 (2010).

\bibitem{resolution} K. W. C. Chan, M. N. O'Sullivan, and R. W. Boyd, \emph{Optimization of thermal ghost imaging: high-order correlations vs. background subtraction}, Opt. Exp. \textbf{18,} 5562 (2010).

\end{thebibliography}
\end{document}